\documentclass[a4paper,11pt]{article}
\usepackage{pos}

\title{Bottomonia screening masses from 2 + 1 flavor QCD}

\author[a]{Peter Petreczky}
\author*[b]{Sayantan Sharma}
\author[c]{Johannes Heinrich Weber}

\affiliation[a]{Physics Department, Brookhaven National Laboratory, \\ 
Upton NY 11973, USA}

\affiliation[b]{The Institute of Mathematical Sciences, HBNI,\\
 Chennai, 600113, India}

\affiliation[c]{Institut f\"ur Physik, Humboldt-Universit\"at zu 
Berlin \& IRIS Adlershof, \\ D-12489 Berlin, Germany}

\emailAdd{sayantans@imsc.res.in}

\abstract{The sequential melting of the bottomonium states is one of the important 
signals for the existence of a quark gluon plasma. The study of bottomonia spectral 
functions on the lattice is a difficult task for many reasons. Calculations based on NRQCD, 
that are commonly used for such purpose, are not applicable at high temperatures. In this 
work we propose a new method to study this problem by calculating the spatial screening 
masses of bottomonium states. We calculate the spatial meson correlators and extract the 
screening masses for mesons in different quantum channels using highly improved staggered 
quark (HISQ) action for bottom quarks and dynamical $2+1$ flavor QCD HISQ gauge configurations. 
The typical lattices we choose are of size $N_s^3 \times N_\tau$ where $N_s=4 N_\tau$ and 
$N_\tau=8, 10, 12$. We consider the temperature range $T = 300$-$1000$ MeV. We show that 
for $T > 500$ MeV the temperature dependence of the screening masses of the ground state 
bottomonia are compatible with the expectations based on uncorrelated quark anti-quark pairs.}

\FullConference{%
 The 38th International Symposium on Lattice Field Theory, LATTICE2021
  26th-30th July, 2021
  Zoom/Gather@Massachusetts Institute of Technology
}

\begin{document}
\maketitle

\section{Introduction}

The dissolution of the bound state of heavy quarks called 
quarkonium has been suggested as  one of the remarkable signals of the formation 
of a quark-gluon plasma in heavy-ion collisions~\cite{Matsui:1986dk}.  
Recent experimental results from CMS have given some tantalizing signals
of the relative suppression of bottomonium  $\Upsilon(2s)$ state 
compared to its ground state~\cite{CMS:2016rpc} and further experimental 
efforts are ongoing to understand charmonium suppression at RHIC and LHC. 
At the same time,  recent phenomenological studies of the dynamics of 
quarkonia have produced many new insights about the widths and lifetimes 
of such states, see for example, Refs.~\cite{Aarts:2016hap,Mocsy:2013syh} 
for recent reviews.

In principle, the properties of quarkonium are all encoded in its spectral function
defined in terms of the real-time correlation function of the appropriate hadron 
operator. At finite temperature, the peaks in the spectral function at frequencies 
equal to the masses of the bound states  are expected to be broadened and shifted in 
the frequency domain, and eventually merge into the continua of quark-antiquark
scattering states. However calculating the spectral functions non-perturbatively 
in a quantum field theory is immensely challenging. Within 
lattice QCD it is possible, in-principle, to calculate the quarkonium
spectral functions through an analytic continuation of the Euclidean 
time correlation functions for that specific quantum number channel, 
calculated using lattice field theory techniques. However, the 
reconstruction of the spectral function from a discrete set of 
Euclidean correlator data is an ill-defined problem. A specific 
reconstruction method of the spectral functions using the maximum 
entropy method has been used extensively and the initial results 
on quarkonia spectral functions can be found 
in~\cite{Nakahara:1999vy,Asakawa:2003re,Karsch:2002wv,
Umeda:2002vr,Datta:2003ww,Jakovac:2006sf}.  At higher 
temperatures, limited extent of the lattice along the temporal 
direction adds to the systematic difficulties in  extracting the in-medium 
quarkonium functions~\cite{Mocsy:2007yj,Petreczky:2008px}.

For bottomonium an additional systematic uncertainty in the lattice results 
arise due large bottom quark mass in the form of mass-dependent discretization
errors. Attempts to circumvent this problem using non-relativistic QCD (NRQCD), 
have yielded several interesting insights on bottomonium properties~
\cite{Lepage:1992tx,Davies:1994mp,Meinel:2010pv,Hammant:2011bt,Dowdall:2011wh}.
Recent lattice studies within the NRQCD effective 
theory~\cite{Aarts:2010ek,Aarts:2014cda,Kim:2014iga,Kim:2018yhk,Larsen:2019bwy,Larsen:2019zqv} 
have shown that ground states of bottomonium channels, 
$\Upsilon(1S)$ and $\eta_b(1S)$ can survive up to temperatures of 
$400$ MeV, whereas significant disagreements exist over the fate of 
$P$-wave bottomonium in the quark-gluon plasma. Furthermore, since 
NRQCD is a long-distance effective theory which is defined only for 
not too small lattice spacings,  we cannot study the high temperature 
regime as $T=1/(a N_{\tau})$ with $N_{\tau}$ being the temporal extent. 
We have thus proposed to study the problem of bottomonia melting using a 
new observable, the screening mass corresponding to different bottomonia 
channels in Ref.~\cite{Petreczky:2021zmz}, the results of which we 
highlight in this talk.

\section{Bottomonium screening correlation functions on the lattice}

The spatial correlation functions can offer a different perspective on the problem 
of in-medium modification of mesons since these are in turn related to meson 
spectral functions at non-zero momenta through the relation,
\begin{equation}
G(z,T)=\int_{0}^{\infty} \frac{2 d \omega}{\omega} \int_{-\infty}^{\infty} 
d p_z e^{i p_z z} \sigma(\omega,p_z,T).
\label{eq:spatial}
\end{equation}
At large distances the spatial meson correlation function decays exponentially, 
and this exponential decay is governed by the \emph{screening mass}, extracted 
from the relation $G(z,T) \sim \rm{exp}(-M_{scr}(T)~ z)$. For  well-defined 
bound state peaks in the meson spectral functions, the corresponding screening 
mass is simply the pole mass. When the quark and anti-quarks are eventually 
unbound at high temperatures, the screening mass is given by 
$2\sqrt{(\pi T)^2+m_q^2}$, with $m_q$ being the quark mass. Thus the 
temperature dependence of the meson screening masses can provide some 
valuable information about the melting of meson states.  Moreover the 
spatial meson correlation functions can be calculated for large spatial
separations between the quark and anti-quarks,  therefore are more sensitive 
to in-medium modifications of meson states~\cite{Karsch:2012na,Bazavov:2014cta}.

We study the bottomonia screening masses on the lattice using the 
highly improved staggered quark (HISQ) discretization for the quarks. 
Not only does the choice of HISQ action minimize the taste-splitting 
for the light quark states, it also preserves the correct dispersion 
relation for heavy quarks ~\cite{Follana:2006rc}.
Within the staggered fermion formalism the meson currents are defined 
as 
\begin{equation}
J_M(\mathbf{x})=\bar q(\mathbf{x}) (\Gamma_D \times \Gamma_F) q(\mathbf{x}), ~\mathbf{x}=(x,y,z,\tau),
\end{equation}
where $\Gamma_D, \Gamma_F$ are the Dirac gamma-matrices corresponding 
to the spin and the staggered tastes. We will consider only the case 
where $\Gamma_D=\Gamma_F=\Gamma$. This corresponds to local operators 
for the meson currents, which in terms of the staggered quark fields 
$\chi(\mathbf{x})$ and the phases $\tilde \phi(\mathbf{x})$ have a 
simple form $J_M(\mathbf{x})=\tilde \phi(\mathbf{x}) \bar \chi(\mathbf{x})
\chi(\mathbf{x})$. Specific choices of the staggered phases correspond to 
mesonic excitations with different quantum numbers. Summing over the temporal 
and any two other spatial coordinates, we arrive at the expression of the 
screening correlator  
$C_M(z)=\sum_{x,y,\tau} \langle J_M(\mathbf{x}) J_M(0)\rangle$.
The staggered meson correlation function contains contributions from 
both parity states which correspond to the oscillating (O) and 
non-oscillating (NO) parts in the correlator. For the lowest 
energy states the screening correlation function for the mesons 
can be simply written as
\begin{eqnarray}
\nonumber
 C_M(z)&=&A_{NO} \cosh\left[M_{NO}\left(z-\frac{N_s}{2}\right)\right] -(-1)^z 
 A_{O} \cosh\left[M_O+\left(z-\frac{N_s}{2}\right)\right].
\label{fit}
\end{eqnarray}
For example, the NO part of the correlator for the staggered 
phase $\tilde{\phi}=-1$ correspond to the pseudo-scalar $\eta_b$ 
channel whereas the corresponding oscillating part correspond to 
the scalar $\chi_{b0}$ excitation. For the quantum number assignments 
to different bottomonium states, see Ref.~\cite{Petreczky:2021zmz}.

\section{Numerical setup}
We calculate the screening masses of the bottomonium states in QCD with 
$2+1$ flavors of dynamical light and strange quarks and treat the 
bottom quarks in the quenched approximation. We have used the dynamical 
gauge configurations which were generated  by the HotQCD 
collaboration~\cite{Bazavov:2014pvz,Ding:2015fca} 
using tree-level Symanzik improved gauge action and HISQ action for 
the quarks. The strange quark mass, $m_s$ was chosen close to its 
physical value, while the light quark masses $m_l=m_s/20$ correspond 
to a Goldstone pion mass of $160$ MeV in the continuum limit~\cite{Bazavov:2014pvz}. 
We performed our study on a wide temperature range between $2-7~T_c$, 
where the chiral crossover temperature $T_c=156.5(1.5)$ MeV ~\cite{HotQCD:2018pds}. 
This was done in order to understand the entire thermal evolution 
of the bottomonium screening correlators. In order to estimate the 
cut-off dependence of our results at each temperature, we considered 
three different temporal extents of $N_\tau=8,10,12$ but the spatial 
extent fixed in each case to be $N_s=4 N_\tau$. With these 
choices of lattice spacings we have ensured that $m_b a \leq 1$ 
for the entire temperature range such that the lattice artifacts 
are sufficiently small in the  bottomonium correlators.

The bottom quark mass in this entire range has been set to be 
$52.5m_s$, which is very close to its physical value. The details 
of the lattice parameters including the inverse of the bare lattice 
gauge coupling $\beta=10/g_0^2$, quark masses, temperatures as well 
as the number of configurations used are mentioned in Table~\ref{tab:latpar}. 
The values of temperature  were set using  $r_1$-scale where we 
have taken $r_1=0.3106(18)$ fm~\cite{MILC:2010hzw}. 
We have used point sources corresponding to both the bottom quark 
and its  anti-particle in the meson correlators and performed two-state 
fits of the corresponding correlators using Eq. (\ref{fit}) in order to 
determine the bottomonium screening masses. 

\begin{center}
\begin{table}[h]
\begin{tabular}{|l|l|l|l|c|l|c|l|c|}
\hline
$\beta$ & $a m_s$ & $am_b$ & \multicolumn{2}{|c|}{$N_\tau=8$}  & \multicolumn{2}{c|}{$N_\tau=10$} & \multicolumn{2}{c|}{$N_\tau=12$} \\ 
\hline
      &       &      & $T$ (MeV)      &\# confs    & $T$ (MeV)       &\# confs       & $T$ (MeV)       &  \# confs     \\
\hline 
7.650 & 0.0192& 1.01 & -         & -      & -         &  -        & 349       &   220	  \\	
\hline
7.825 & 0.0164& 0.86 & 610	 & 500	  &  488      &  250	  & 407       &   180	  \\	
\hline
8.000 & 0.0140& 0.74 & 710       & 500	  & 568       &	500	  & 473	      &   180	\\	
\hline
8.200 & 0.0117& 0.61 & 842       & 250	  & 674       &	250	  & 561       &   500	\\	
\hline
8.400 & 0.0098& 0.52 & 998       & 240	  & 798       &	250	  & 665       &   500	\\	
\hline
8.570 & 0.0084& 0.44 & -         & -      & 922       &	250	  & 768       &   250	\\	
\hline
8.710 & 0.0074& 0.39 & -         & -      & -         & -         & 864       &   250	\\	
\hline
8.850 & 0.0065& 0.34 & -         & -	  & -         & -         & 972       &   250	\\	
\hline
\end{tabular} 
    \caption{The entire list of inverse bare gauge couplings, $\beta$, quark masses, 
    temperature values and the corresponding number of gauge configurations 
    (\#confs) used in this study.}
    \label{tab:latpar}
\end{table}
\end{center}

\section{Results} 
We summarize here our main findings, for a more detailed discussion 
of our results for bottomonia screening masses we refer to our published
work~\cite{Petreczky:2021zmz}. 

\begin{figure}
\includegraphics[scale=0.7]{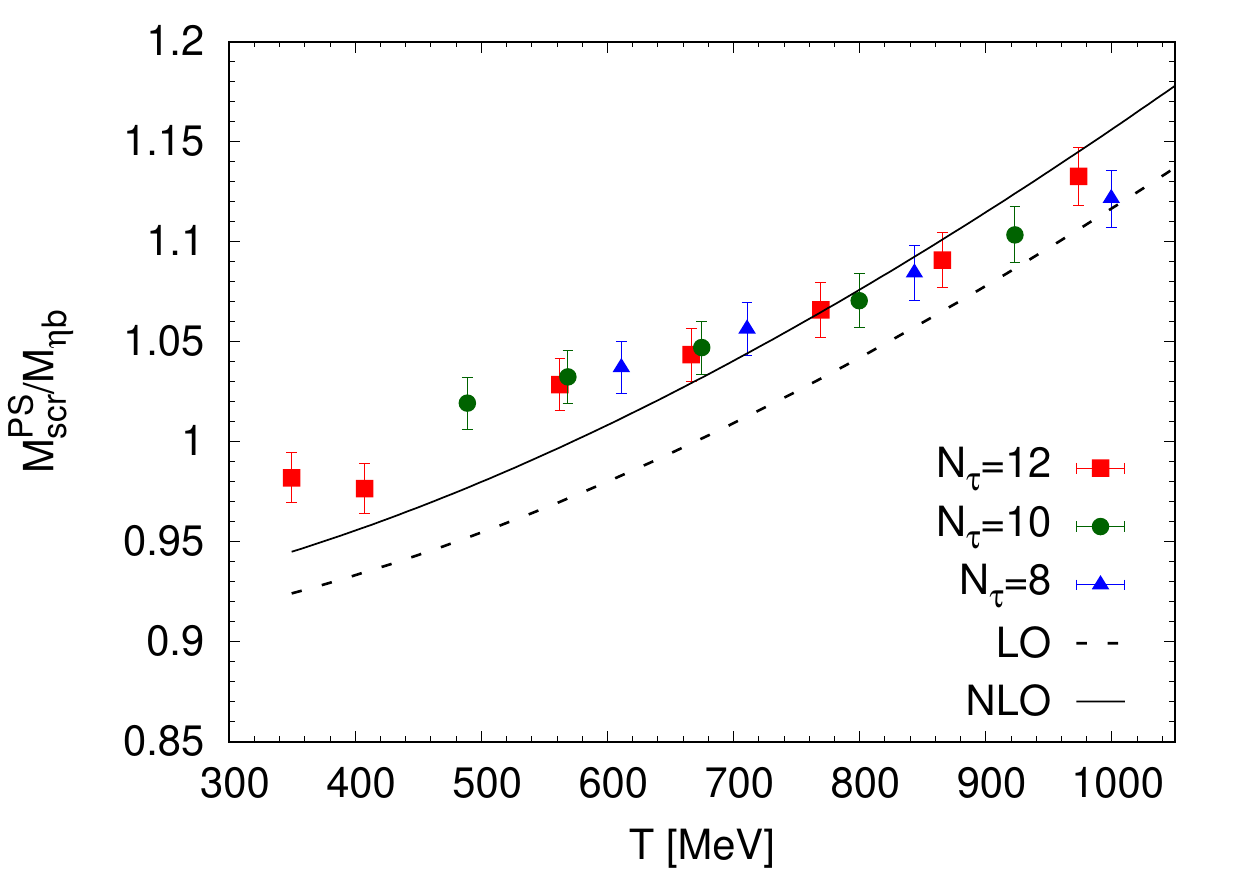}
\caption{The pseudo-scalar screening mass normalized 
by the mass of $\eta_b(1S)$ meson at zero temperature,
as function of the temperature obtained on lattices with 
$N_{\tau}=8,~10$ and $12$. The solid line depicts the 
prediction for the screening mass at lowest order (LO) 
in perturbation theory, while the dashed line is the NLO 
prediction, taken from Ref.~\cite{Petreczky:2021zmz}.}
\label{fig:PS}
\end{figure}

\begin{figure}
\includegraphics[scale=0.6]{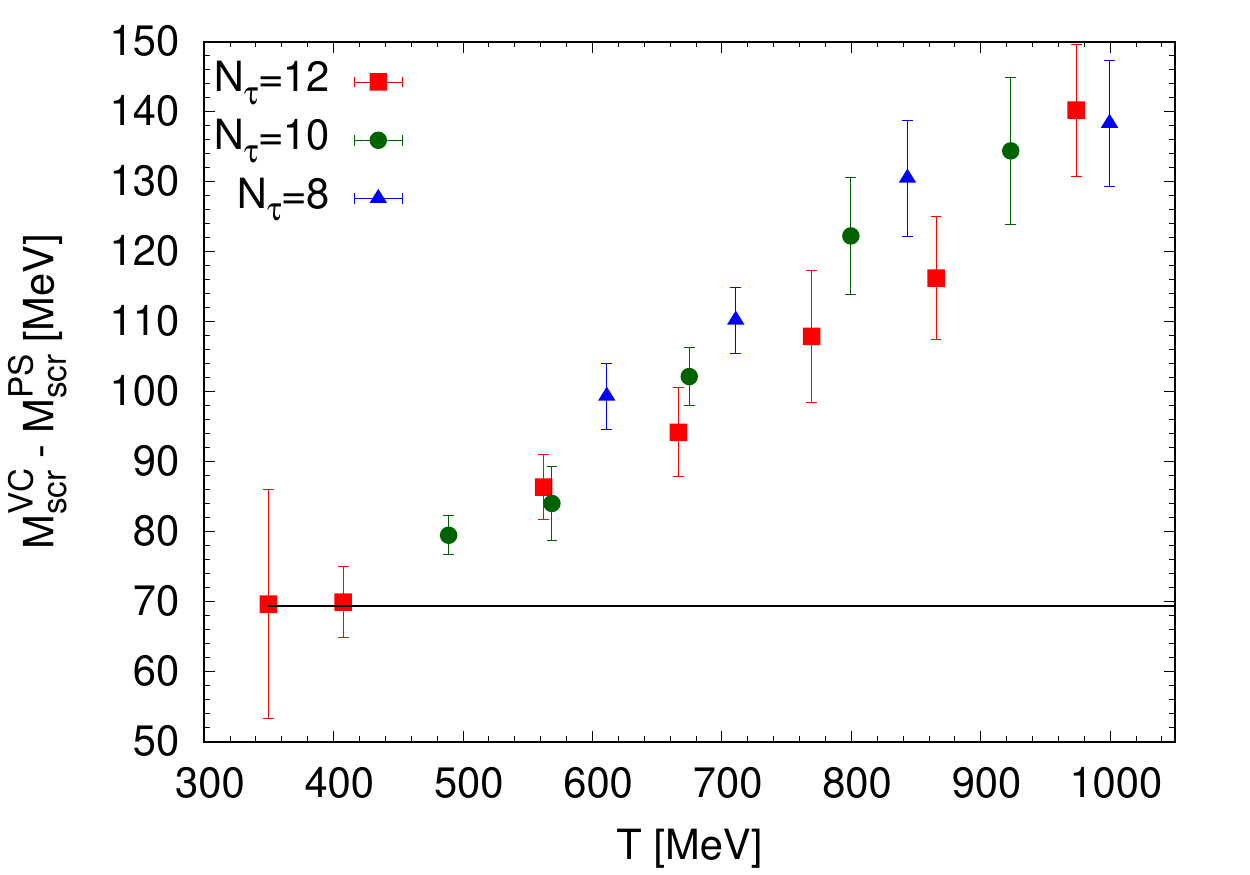}
\includegraphics[scale=0.6]{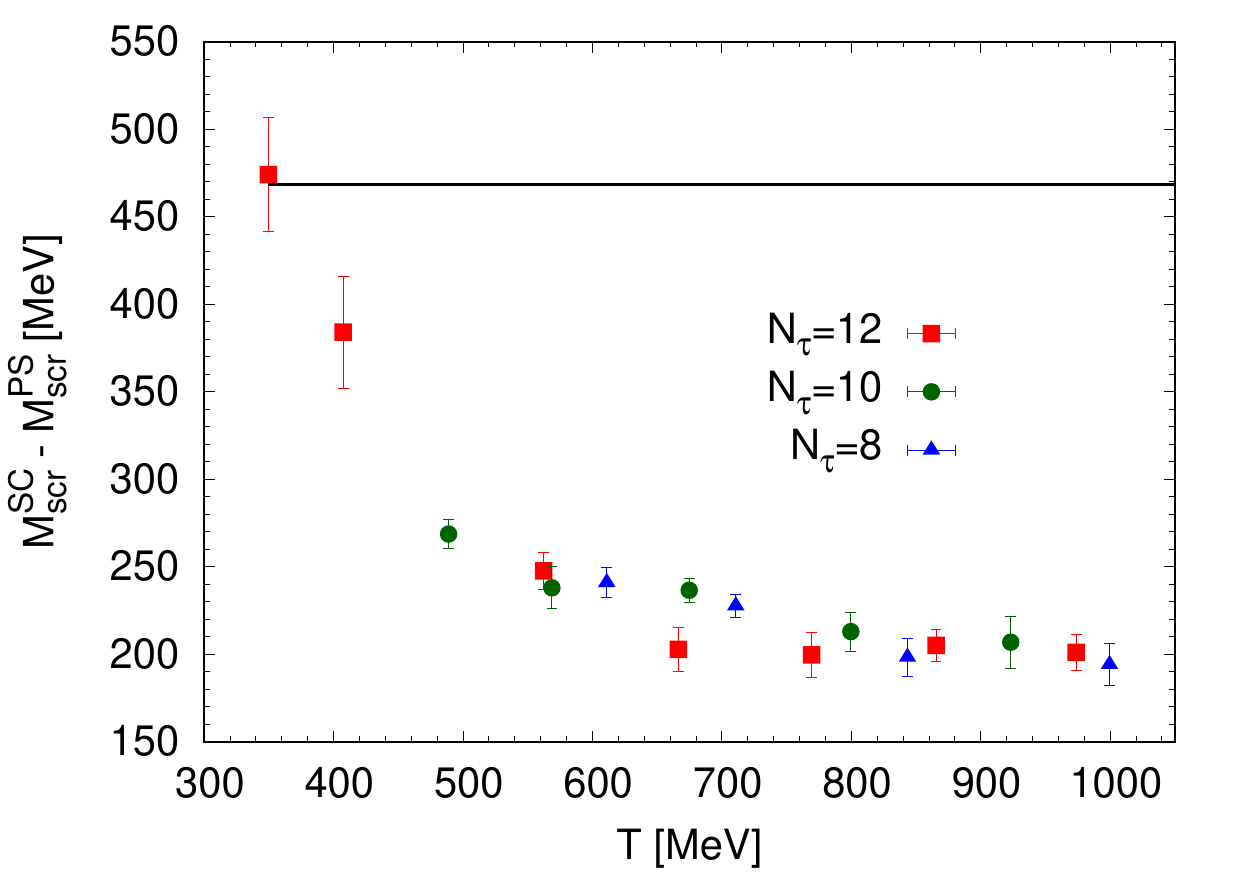}
\caption{The left panel shows the difference between the vector 
and pseudo-scalar screening masses as function of the temperature 
obtained on lattices with $N_{\tau}=8,~10$ and $12$, from Ref.~\cite{Petreczky:2021zmz}. 
The solid line corresponds to the difference between the 
$\Upsilon(1S)$ mass and the $\eta_b(1S)$ mass reported in 
the Particle Data Group. The right panel shows the difference 
between the scalar and pseudo-scalar screening masses as 
function of the temperature obtained on lattices with 
$N_{\tau}=8,~10$ and $12$, from Ref.~\cite{Petreczky:2021zmz}. The 
solid line corresponds to the difference between the $\eta_b(1S)$ and 
the $\chi_{b0}(1P)$ mass, from Particle Data Group.}
\label{fig:VC-PS}
\end{figure}

\begin{itemize}
\item 
At the lowest temperatures, the $\eta_b$ screening mass is 
close to the zero temperature mass, while at high temperatures 
$T>500$ MeV the screening mass increases linearly with temperature, 
signalling melting of $\eta_b$ as shown in Fig. \ref{fig:PS}.

\item
Also seen in  Fig. \ref{fig:PS}, the ratio 
$M^{\text{PS}}_{\text{scr}}/M_{\eta_b}$ 
shows a slow convergence to its corresponding 
values from NLO perturbation theory at 
$T>500$ MeV, suggesting a gradual melting of 
$\eta_b$ into a pair of correlated $b$-$\bar{b}$ pairs.

\item 
As evident from the left panel of Fig.~\ref{fig:VC-PS} at the lowest two 
temperatures the difference between the vector and 
pseudo-scalar screening masses are consistent with 
the hyperfine splitting that exists between the 
$\Upsilon(1S)$ and $\eta_b(1S)$ states which is 
about $70$ MeV~\cite{ParticleDataGroup:2018ovx}. 
This suggests that at these temperatures the $\eta_b(1S)$ 
and $\Upsilon(1S)$ exist as well defined bound states with 
little in-medium modifications.

\item 
For $T>450$ MeV the difference between vector and 
pseudo-scalar screening masses increases linearly 
with temperature as shown in the left panel of Fig.~\ref{fig:VC-PS}.  
Within perturbation theory at NLO in strong-coupling 
constant this difference is identically zero.  
One can understand this observation within a three 
dimensional effective theory of QCD. In this effective 
theory, a quark and anti-quark propagating along the 
$z$-direction, interact via a spin-dependent potential 
which is proportional to the temperature~\cite{Koch:1992nx} 
giving a mass splitting of $0.3~T$ at $T>900$ MeV. 

\item
Since these spin-dependent interactions are inversely 
proportional to the quark mass, its effects in the mass 
splitting between vector and pseudo-scalar screening 
states is negligibly small for the bottom sector as 
compared to the light quark sector. 

\item 
Furthermore from the right panel of Fig.~\ref{fig:VC-PS}, the difference between 
the scalar and pseudo-scalar (or between the axial-vector and 
vector) screening masses  agree with the  differences between 
the $\chi_{b0}(1P)$ and $\eta_b(1S)$ ($\chi_{b1}(1P)$ and 
$\Upsilon(1S)$) masses reported in the Particle Data Group
~\cite{ParticleDataGroup:2018ovx}. This again suggests 
that $\chi_{b0}(1P)$ and $\chi_{b1}(1P)$ states exist as 
confined states at $T=350$ MeV with relatively small medium 
modifications. 
A large drop in this observable is seen for $350<T<600$ MeV, 
which approaches zero gradually owing to tiny effects of the 
bottom quark mass. Eventually in the regime when $T\gg m_b$, 
this difference should vanish.
\end{itemize}

\section{Discussion on the possible systematics}

\begin{itemize}

\item 
\emph{Setting the bottom quark mass:}
We have chosen the ratio of $m_b/m_s=52.5$ which is close to its 
physical value. However the lines of constant physics for the 
strange quark mass have not been fixed very precisely 
for the entire temperature range we have studied~\cite{Bazavov:2014pvz}. 
Hence it is not possible to simply set the $\eta_b$ mass 
(used in the ratio shown in Fig.~\ref{fig:PS}) to its 
experimentally measured value.  From  the estimated dependence 
of the $\eta_b$ mass on the $b$ quark mass for the HISQ
action~\cite{Petreczky:2019ozv}, it turns out that the $\eta_b$ 
mass is larger than the PDG value by $4\%$ and $9.8\%$ for $\beta=7.596$ 
and $\beta=8.4$ respectively. At still higher values of $\beta$ where 
no previous measurements exist, we assume that the $\eta_b$ mass is 
$9.8\%$ larger than the experimentally measured value.

\item
\emph{Sources of systematic errors in the data: }
We did not calculate the zero temperature mass for $\eta_b$ 
explicitly for all the lattice spacings we have studied but 
estimated them based on interpolation, thus we have assigned a 
systematic error of $1\%$ to the zero temperature mass for 
$\eta_b$. Moreover the errors in the scale setting of $a/r_1$ 
as well as the error in measuring $r_1$ in physical units also 
adds to the systematic error. All these error estimates were 
added in quadrature and combined together with the statistical 
errors, in order to estimate the error bars in the data points of 
Fig.~\ref{fig:PS}. In the screening mass differences shown in 
Fig.~\ref{fig:VC-PS}, the effects due to 
the small deviation of bottom quark mass from its physical value 
is tiny. Thus in estimating the errors for these observables, we 
have simply added  the errors in the determination of lattice 
spacings and the statistical errors in quadrature. 

\item
\emph{Cut-off effects in the data: }Quite surprisingly and assuredly 
we do not observe significant cut-off dependence in the screening mass 
results in the bottomonium sector, since the data agrees within errors 
for all three different $N_\tau$ values in all the plots.

\end{itemize}

\section{Summary and outlook}
We have presented here the main findings of a first comprehensive study 
about the temperature dependence of the bottomonium screening masses on 
the lattice using a relativistic action (HISQ) for the bottom quarks. The
observables we have proposed are rather insensitive to finite lattice 
cut-off effects, allowing us to make some very robust 
predictions~\cite{Petreczky:2021zmz}. To summarize our main findings,

\begin{itemize}
\item 
Our estimate of $\eta_b$ melting temperature is about $450$ MeV. 
This is consistent with the results from NRQCD based studies~\cite{Aarts:2010ek,Aarts:2014cda,Kim:2018yhk,Larsen:2019bwy,Larsen:2019zqv} 
as well as with the results from potential models with a screened 
complex potential~\cite{Petreczky:2010tk,Burnier:2015tda}.

\item
We observe a linear dependence of the $\eta_b$ mass with temperature 
for $T>500$ MeV. For charmonium the linear increase with the temperature 
for $\eta_c$ screening mass is seen already at $T>250$ MeV~\cite{Bazavov:2014cta} 
showing a clear mass-hierarchy in the melting of quarkonia. 

\item 
The $1P$-bottomonia melt already at a comparatively lower temperature 
$\sim 350 $ MeV, thus showing that the ground state bottomonia are more 
robust against in-medium modifications compared to their excited states.

\end{itemize}

\textbf{Acknowledgments:}
PP was supported by U.S. Department of Energy under Contract No. DE-SC0012704. 
JHW work was supported by U.S. Department of Energy, Office of Science, 
Office of Nuclear Physics and Office of Advanced Scientific Computing Research within 
the framework of Scientific Discovery through Advance Computing (SciDAC) award 
Computing the Properties of Matter with Leadership Computing Resources, and by 
the Deutsche Forschungsgemeinschaft (DFG, German Research Foundation) - Projektnummer 
417533893/GRK2575 ``Rethinking Quantum Field Theory''. SS gratefully acknowledges 
financial support from the Department of Science and Technology, Govt. of India, through a 
Ramanujan Fellowship. 

\bibliographystyle{JHEP}
\bibliography{references_ssharma.bib}

\end{document}